\documentclass[aps,prx,twocolumn,amsmath,amssymb,longbibliography,superscriptaddress]{revtex4-1}

\usepackage{graphicx}
\usepackage{dcolumn}
\usepackage{bm}
\usepackage{multirow}
\usepackage[utf8]{inputenc}

\usepackage{color}

\usepackage[dvipsnames]{xcolor}

\begin{document}

\title{\textit{Vector-chirality} driven topological phase transitions in noncollinear antiferromagnets and its impact on anomalous Hall effect} 
\author{Subhadip Pradhan}
\affiliation{School of Physical Sciences, National Institute of Science Education and Research, An OCC of Homi Bhabha National Institute, Jatni-752050, India}
\author{Kartik Samanta}
\affiliation{PGI-1 and IAS-1, Forschungszentrum J\"ulich and JARA, 52425 J\"ulich, Germany}
\author{Kush Saha}%
\affiliation{School of Physical Sciences, National Institute of Science Education and Research, An OCC of Homi Bhabha National Institute, Jatni-752050, India}
\author{Ashis K.\ Nandy}
\email{aknandy@niser.ac.in}
\affiliation{School of Physical Sciences, National Institute of Science Education and Research, An OCC of Homi Bhabha National Institute, Jatni-752050, India}




\maketitle
\section*{abstract}
Magnetic materials showing topologically nontrivial quantum states with high tunability is an undoubtedly important topic in condensed matter physics and material science. Based on the first-principles electronic structure calculations and subsequent symmetry adapted effective low-energy $\textbf{k.p}$ theory, we show in a noncollinear antiferromagnet (AFM), Mn$_3$Sn, that the switching of the \textit{vector-chirality}, $\kappa$, is an unconventional route to topological phase transition from a nodal-ring to a Weyl point semimetal. Specifically, we find that the switching of $\kappa$ via $staggered$ rotation leads to gapping out an elliptic nodal-ring everywhere at the Fermi-level except for a pair of points on the ring. As a consequence, the topological phase transition switches the anomalous Hall conductivity (AHC) from zero to a giant value. Furthermore, we theoretically demonstrate how the controlled manipulation of the chiral AFM order keeping $\kappa$ unaltered favors unusual rotation of Weyl-points on the ring. In fact, without $staggered$ rotation, this enables us to tune and switch the sign of in-plane components of the AHC by a collective uniform rotations of spins in the AFM unit cell.

\section{Introduction}
Over the past decade, a variety of topological semimetals (TSMs) \cite{RevModPhys.90.015001,doi:10.1146/annurev-matsci-070218-010049,Hasan_2021,Yang_2016}, namely, nodal point semimetals \cite{PhysRevLett.108.140405,PhysRevB.83.205101,PhysRevB.99.075116}, nodal line/ring semimetals \cite{PhysRevB.92.045108,Fang_2016,PhysRevLett.122.077203,Belopolski_2019,PhysRevB.104.235136}, nodal surface semimetals \cite{PhysRevB.97.075120,D0TC01978J,PhysRevB.97.235150} have taken widespread attention due to the unusual transport phenomena related to the protected nontrivial boundary states. 
Most nodal point semimetals fall into two categories: Weyl semimetals (WSMs) \cite{huang2015weyl,Yang_2017,Kuroda_2017,Yan_2017} and Dirac semimetals (DSMs) \cite{Liu_2014,doi:10.1021/jacs.8b09900,C3CP53257G,hosen2018distinct}, depending on the degeneracies and distribution of the band crossing nodes in the bulk band structures. A four-fold degenerate Dirac node can be thought of a point with a degenerate pair of Weyl points (WPs) carrying opposite chiral charges, where WPs can be derived by separating Dirac points in momentum space by breaking either time reversal symmetry or inversion symmetry.
The nodal-ring semimetals (NRSMs) with two fold degenerate one-dimensional band crossings on the other hand can be gapped out except a pair of discrete touching points, an alternative approach to create WPs. The advantage here is the controlled generation of the discrete touching pairs on a loop. 
Recently, coexistence of magnetism, in particular ferromagnetism and topological quantum states in band structures has emerged as an important platform both for fundamental and technological interests \cite{2019NatRP...1..126T,naturematerial}.
In conventional topological ferromagnets, the ability to externally control the magnetism by magnetic, electrical and optical fields offers a remarkable way of transitions between topological phases \cite{PhysRevLett.115.036805,zhang2014electrically,doi:10.1073/pnas.1713458114,article}. 
Subsequently, attention has been drawn to nontrivial topology in magnetic materials without a net magnetization \textit{i.e.} antiferromagnets (AFMs) \cite{PhysRevX.9.041040,Otrokov_2019,wangC,liuH}.
Note, the recent developments of electrical manipulation and detection of AFM orders \cite{Wadley_2016,Zelezny}, particularly in view of topological AFM spintronics \cite{Nature_anti,prl118,diang-Fu2019} open up possibilities to find topological phase transitions in AFMs too. 
This is due to the fact that the current induced $staggered$ torque \cite{salemi2019orbitally} allows unprecedented control on AFM orders which, in turn, may change symmetries of AFMs significantly to host various topological quantum phases.
Nonetheless, no noncollinear AFM has been realized to show topological phase transitions under the change in the noncollinear order and possibly, we need an unconventional route to manipulate the AFM order.
\begin{figure*}[ht!]
\begin{center}
      \includegraphics[width=\textwidth]{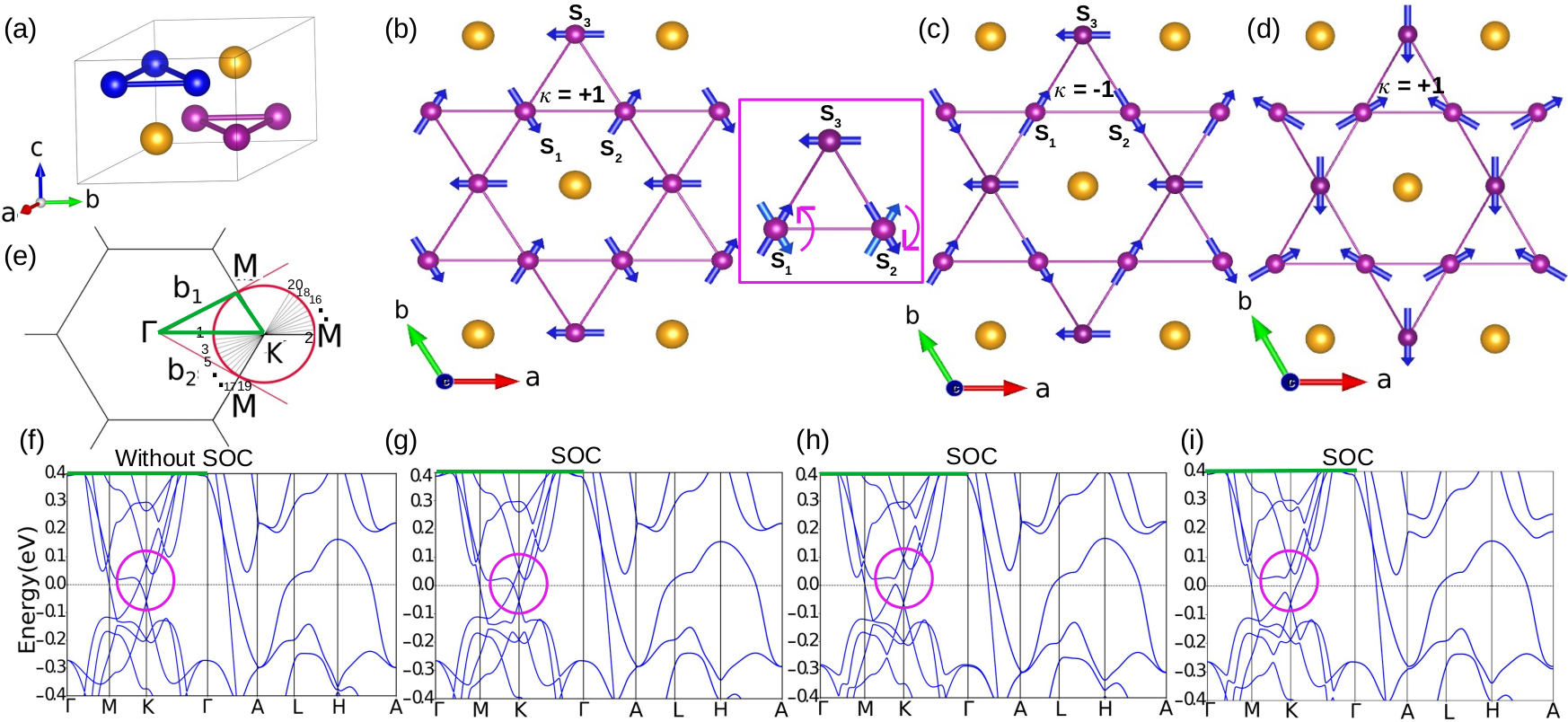}
      \caption{\textbf{Noncollinear antiferromagnetic structures with non-zero vector chirality and the corresponding electronic structure of Mn$_3$Sn.}~{\bf (a)} Mn$_3$Sn bulk crystal structure, blue and magenta balls represent the Mn atoms in top and bottom magnetic kagome layers at $z=1/4$ and $3/4$, respectively. The yellow balls represent the Sn atoms. The noncollinear chiral antiferromagnetic textures are presented in one kagome layer, named {\bf (b)} ``direct", {\bf (c)} ``inverse" and {\bf (d)} \textit{3-in-3-out} magnetic states with their corresponding $\kappa$ values. The square box here depicts how switching of $\kappa$ can be achieved via simultaneous $staggered$ rotations of \textbf{S}$_1$ and \textbf{S}$_2$. {\bf (e)} The 2D projection of the hexagonal Brillouin zone (BZ) of Mn$_3$Sn. Essential high symmetry points are marked here. {\bf (f)} The band structures without considering the spin-orbit coupling (SOC). Note, for all 120$^o$ noncollinear states band structures remain identical unless the SOC is switched on. After considering the SOC effect, {\bf (g)}-{\bf (i)} show the band structure for the magnetic states ``direct", ``inverse" and \textit{3-in-3-out}, respectively. The nontrivial features near $E_f$ are highlighted within the red circles. The thick green line in \textbf{(e)} and the corresponding band structure plots \textbf{(f)}-\textbf{(i)} highlights the $k$-path for the $k_z=0$ plane of the BZ.}
      \label{fig:mn3sn_a}
\end{center}
\end{figure*}
The noncollinear AFM orders in Mn based kagome-like planes have been subsequently discovered in a series of Heusler materials, cubic Mn$_3$Y (Y = Rh, Ir, Pt) \cite{PhysRevLett.112.017205,PhysRevB.92.144426} and hexagonal Mn$_3$X (X = Ge, Sn, Ga) \cite{K_bler_2014,nakatsuji2015large,doi:10.1126/sciadv.1501870,doi:10.1063/1.5143959,PhysRevB.95.075128,Ikhlas_2017,csingh,PhysRevLett.76.4963} materials. Since then, such AFM orders become an important topic of research, as they are found to show various TSM phases \cite{Tsai_2020,PhysRevB.99.104428}, however, the noncollinear magnetization driven transitions between various TSM phases are missing. Here, the noncollinear AFM order can be classified based on defining a vector spin chirality\cite{Kawamura_2001} as,
\begin{equation}
    \kappa = \frac{2}{3\sqrt{3}} \sum_{ <ij>}[{\hat{n}_i }\times {\hat{n}_j}]_{z},
    \label{VC}
\end{equation}
where $<\! {ij}\! >$ runs over the nearest neighbor spins, $\textbf{S}_i$ = S \!\!$\hat{n}_i$ with $\hat{n}_i$ as the unit vector.
The magnitude of $\kappa$ is unity while the signs `$+$'  and `$-$' represent the chiral states with magnetic orders, ``direct" and ``inverse" triangular AFM, respectively. Notably, the magnetic ground state of Mn$_3$X and Mn$_3$Y show opposite $\kappa$ value in the kagome sublattices which remains unaltered with simultaneous uniform spin rotations within the plane. However, materials within Mn$_3$Y family are crystallized into a face-centered cubic structure (space group Pm$\overline{3}$m \cite{Zhangprb} and $O_h^1$ magnetic point group \cite{PhysRevB.95.094406}) where magnetic kagome planes are stacked in ``$-ABC-ABC-$" order along the crystallographic (111) axis. This is different in compared to the hexagonal Mn$_3$X family ($D^4_{6h}$ magnetic point group \cite{PhysRevB.95.094406,prb101}) where one finds ``$-AB-AB-$" stacking order along the hexagonal (0001) axis. Therefore, for a fixed $\kappa$, one would expect that the band structures in both family of materials will be significantly different from the symmetry constraints.
In this class of chiral AFMs, $\kappa$ becomes an important quantity which by switching between $+1$ and $-1$ can lead to control various symmetries and hence, offer a possible route to tailor nontrivial topological phases. 

In this work, within a noncollinear chiral AFM, Mn$_3$Sn, we have shown a topological phase transition from NRSM to WSM through an unusual vector chirality, $\kappa$ switching from `+1' to `-1' via introducing a $staggered$ rotation. We note that earlier studies established the Weyl-points (WPs) physics in the ``inverse" triangular AFM state \cite{Yang_2017}. In contrast, the switching to its competing chiral state carrying $\kappa=+1$ (``direct") and exhibiting topological NRSM state is one of the key findings of the present study. The switching of vector chirality $\kappa$  from `+1' to `-1' tunes the anomalous Hall conductivity (AHC) value from zero to a giant value. Moreover, even in the absence of $staggered$ rotation, a simultaneous uniform rotation of Mn moments keeping $\kappa=-1$ serves to rotate the Weyl points and hence, tailor the anomalous Hall effect (AHE). This therefore switches the sign of the in-plane components of the AHC. Finally, we theoretically demonstrate by tuning the chiral AFM order how the TSM phases switch and tailor the nontrivial transport properties.

\section{Results}
To uncover the topological phase transition within the electronic structure theory of noncollinear AFMs, we adopt the density functional theory (DFT) formalism as implemented in the full-potential linearized augmented plane wave (FLAPW) all-electron code, FLEUR \cite{fleur} and the plane-wave projected augmented wave (PAW) pseudopotential code, Vienna ab initio Simulation Package (VASP) \cite{vasp,paw2}. Details of the computational and numerical approaches are provided in the Method subsection.
In the following section, we discuss the interplay between the vector chirality $\kappa$ and $\kappa$ induced nontrivial topological phases in band structures of a noncollinear AFM, Mn$_3$Sn. 
As a consequence, we show how the onset of various topological phases via controlled manipulation of the chiral AFM order drives the system into AHC switching and further tuning of its in-plane components.

\subsection{Noncollinear chiral antiferromagnet: Mn$_3$Sn}
Bulk Mn$_3$Sn, shown in Fig.~\ref{fig:mn3sn_a}(a), crystallizes in a layered hexagonal lattice with space group, $P6_3/mmc$, in which Mn atoms form a magnetic kagome layer, see Fig.~\ref{fig:mn3sn_a}(b), (c), and (d). Each two-dimensional (2D) kagome geometry formed by Mn atoms contains Sn atom at the center of the hexagon of Mn. The following discussions are based on the \textit{ab initio} results calculated with experimental lattice parameters, $a = b = 5.65$ \AA, and $c = 4.522$ \AA. \cite{PhysRevResearch.2.043366}
The ``direct" triangular AFM state in Fig.~\ref{fig:mn3sn_a}(b) is distinguished by the \textit{handedness} of the spin rotations: a counterclockwise $120^o$ rotation turns the spin $\textbf{S}_1$ in one atom into $\textbf{S}_2$ in the neighboring atom. 
The ``inverse" triangular AFM texture in Fig.~\ref{fig:mn3sn_a}(c) on the other hand is the result of clockwise $120^o$ rotation while moving from \textbf{S}$_1$ to \textbf{S}$_2$.
One easy way to switch $\kappa$ from `$+1$' to `$-1$' is by a $staggered$ rotation of any two spins while keeping the third one unaltered in the unit cell. 
Here, the $staggered$ rotation means a simultaneous counterclockwise and clockwise rotation of the in-plane spins, e.g. $\textbf{S}_1$ and $\textbf{S}_2$, respectively with respect to the $z$-axis as depicted in the small square box in Fig.~\ref{fig:mn3sn_a}.
By doing so, the $\mathcal{C}_{3z}$ rotational symmetry, a special symmetry that is protected in $\kappa=+1$ state is broken.
In our magnetic calculations without spin orbit coupling (SOC), both noncollinear states are equivalent; \textit{i.e.}, they are degenerate in total energies.
The calculated magnetic moment of each Mn atom is about $3.12~\mu_{\textrm B}$, confined in the $xy$-plane.
These two magnetic states become inequivalent when SOC is switched on and the coplanar $\kappa=-1$ AFM state is found to be the ground state. 
The stability energy with respect to the competing $\kappa=+1$ state is found to be very small, about $4.5~meV/f.u.$
Moreover, the calculated energy difference as a function of $staggered$ rotation is found to exhibit two inequivalent minima corresponding to the $\kappa =+1$ and $-1$ states (see Supplementary Fig.~ 1 and Supplementary Note 1). This indeed corroborates the breaking of vector chiral symmetry in the presence of SOC. 

\subsection{Topological footprints associated with $\kappa$}
The \textit{ab initio} bulk band structure of Mn$_3$Sn calculated within local density approximation without considering the SOC is shown in Fig.~\ref{fig:mn3sn_a}(f) along the high symmetry directions including the special points marked in the 2D projected Brillouin zone (BZ), Fig.~\ref{fig:mn3sn_a}(e). The band structure remains the same for all 120$^o$ triangular noncollinear states \textrm{i.e.} irrespective of the $\kappa$.
The band structure clearly shows two band crossing points around the $K$ point along the $MK$ and $K\Gamma$ directions.
Generally, in the presence of spin-orbit coupling (SOC), the band crossing points have the possibility of either completely gapped out or decaying into WPs. 
However, for the particular $\kappa =+1$ magnetic state shown in Fig.~\ref{fig:mn3sn_a}(b), the band crossings are impervious to the presence of SOC, see the region near $E_f$, marked by a red circle in Fig.~\ref{fig:mn3sn_a}(g). 
Switching to $\kappa =-1$ magnetic state as in Fig.~\ref{fig:mn3sn_a}(c), the band structure in Fig.~\ref{fig:mn3sn_a}(h) shows one crossing point near $E_f$ along $K \Gamma$ direction. The gapped crossing is lying at $E_f$ along $MK$ direction.
The gapless node around which band dispersion is linear is called the 3D WP, lying slightly above $E_f$ along $\Gamma K$ direction.
Therefore, the change in band structure is related to the $\kappa$ value of the underlying AFM order.
The band structure further changes vigorously with the uniform collective rotation of all Mn moments while keeping $\kappa$ constant.
For a spin configuration in Fig.~\ref{fig:mn3sn_a}(d) named \textit{3-in-3-out} state, the band crossings are all gapped out, see Fig.~\ref{fig:mn3sn_a}(i). 
We note that there exists another AFM state with $\kappa=-1$ where the band crossing remains intact along the direction perpendicular to the $\Gamma KM$ line, see Supplementary Fig.~2 and Supplementary Note 2, indicating a rotation of the band crossing points around the $K$ point in the BZ. 
Thus, it is expected to have interesting nontrivial topological features in the band structures with the magnetic order dynamics in Mn$_3$Sn.

\begin{figure}[b]
    \centering
    \includegraphics[width=8.5cm,height = 6cm]{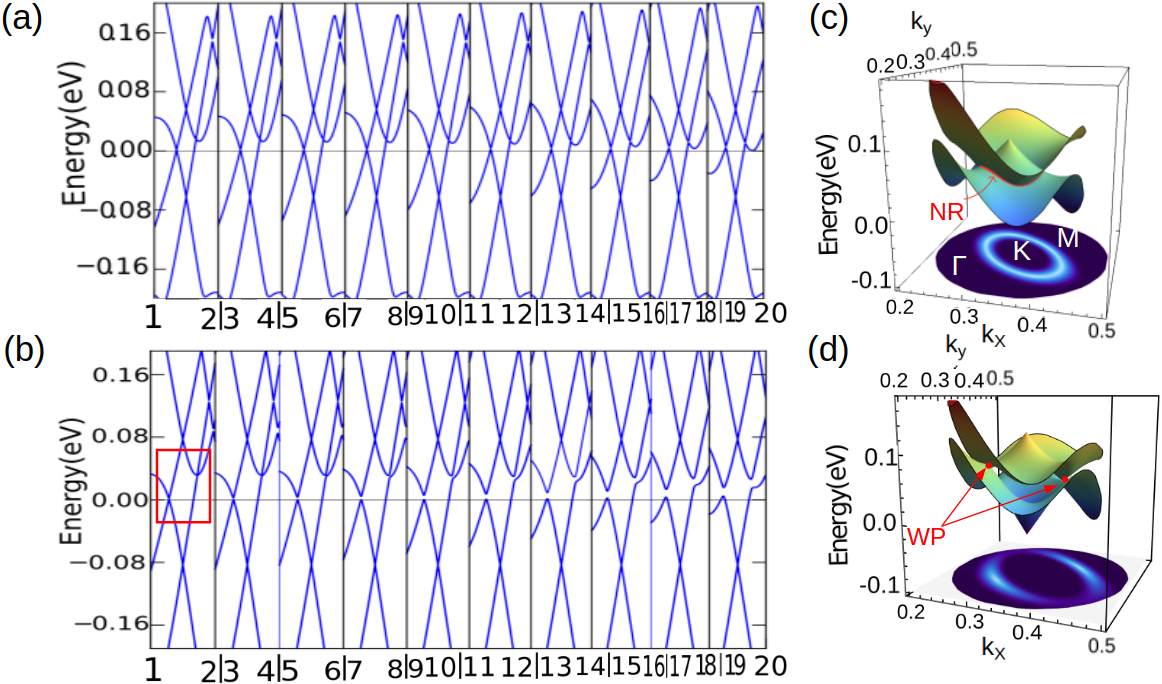}
    \caption{\textbf{Nodal line semimetal and Weyl semimetal phases for different vector chirality ($\kappa$ = `$\pm1$') states.} {\bf (a)} and {\bf (b)}, Band structures calculated with spin-orbit coupling for a particular $\kappa = + 1$ and $-1$ magnetic states, respectively. The $k$-path segments are shown in Fig.~\ref{fig:mn3sn_a}{\bf (b)}. {\bf (c)} and {\bf (d)}, 3D band structures for $\kappa = + 1$ and $-1$, respectively calculated within $k_z=0$ plane. The red line drawn along the band crossings defines the nodal-ring state whose projection on a constant energy plane is an ellipse.}
    \label{fig:band_soc}
\end{figure}

\begin{figure*}[ht!]
    \centering
    \includegraphics[width=\textwidth]{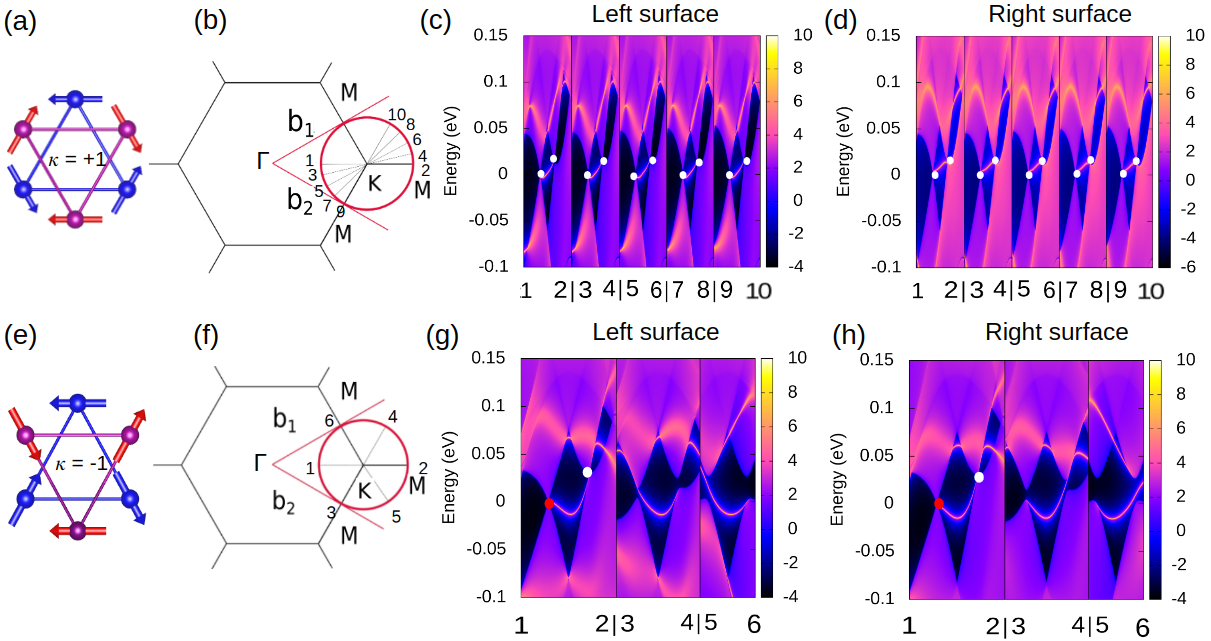}
    \caption{\textbf{Calculated surface states representing the topological drumhead and the Fermi arc.} {\bf(a)} 120$^o$ noncollinear antiferromagnetic configuration carrying vector chirality $\kappa = +1$. {\bf(b)} Different $k$-path segments around K point in the 2D projected Brillouin zone of (001) surface of Mn$_3$Sn.{\bf(c)} and {\bf(d)} are the surface state spectrums along the $k$-path segments in {\bf (b)} for the left and right surfaces, respectively.
    {\bf(e)} and {\bf(f)} are the 120$^o$ noncollinear antiferromagnetic configuration having vector chirality $\kappa = -1$ and its (001) surface Brillouin zone showing different $k$-path segments, respectively. {\bf(g)} and {\bf(h)} show the surface band dispersion for the (001) slab calculated within WannierTools. The red and white dots along ``1-2'' direction denote the projected Weyl-points carrying different topological charges, `-' and `+', respectively. The color bar represents the surface local density of states (LDOS).}
    \label{fig:surface}
\end{figure*}

The bulk band structures of Mn$_3$Sn presented in Fig.~\ref{fig:mn3sn_a}(f)-(i) are complex with a plethora of band crossings. 
However, it is evident that the intrinsic AHE in these systems is the result of non-zero Berry curvature related with magnetic monopoles, WPs, near the $E_f$ \cite{chen2021anomalous}.
We hence focus on the band crossings around the high symmetry $K$ points in the BZ, in an energy window ($\pm~0.1~eV$) around the $E_f$. Note, the band crossings in Fig.~\ref{fig:mn3sn_a}(f) around $K$ point within the energy window $-0.2$ to $-0.14$ eV are all gapped out in the presence of SOC, see Supplementary Note 5 for more details.
The energy dispersions with very high precision, considering several path segments within a circle centered at $K$ point as indicated by the radial black lines inside the red circle in Fig.~\ref{fig:mn3sn_a}(e), are presented in Fig.~\ref{fig:band_soc}(a). 
The crossing points without any gap are clearly visible in the band dispersion plots and all these gapless points together form a nodal-ring (NR) in the $k_z$ = 0 plane, as shown by marking a red line in the three-dimensional (3D) energy band dispersion in Fig.~\ref{fig:band_soc}(c). 
A finite energy variation along the ring is observed and its projection on the constant energy surface shows an ellipse with its minor axis along the $\Gamma K M$ direction.
Hence, this particular magnetic state in Fig.~\ref{fig:mn3sn_a}(b) is an unusual example of a noncollinear AFM where the NRSM phase is protected. 
Next, moving to the $\kappa =-1$ chiral AFM configuration (Fig.~\ref{fig:mn3sn_a}(c)), we summarize the electronic band structures in Fig.~\ref{fig:band_soc}(b), calculated along the same path segments as stated earlier.
The band crossing points are completely gapped out along all $k$-paths except the $k$-path segment, ``1-2", along $\Gamma KM$ direction, as indicated in the red box.
The gaps are clearly visible for other segments, as an example, see $k$-path segment ``19-20".
We have identified a pair of bulk WPs as presented in the 3D energy band dispersion plot in Fig.~\ref{fig:band_soc}(d).

To elucidate further, topologically nontrivial surface states corresponding to different $\kappa$ values are calculated based on the tight-binding model which is constructed with the maximally localized Wannier functions \cite{MOSTOFI20142309} within WannierTools software package \cite{2018CoPhC.224..405W}.
The calculated results are presented in Fig~\ref{fig:surface}. 
The NR state is, in general, confirmed by the drumhead-like surface states in the surface calculations and hence, one can find surface bands across any pair of diametrically opposite points on the NR.
In case of a particular AFM state carrying $\kappa=+1$ (Fig.~\ref{fig:surface}(a)), Fig.~\ref{fig:surface}(c) and (d)  show the surface dispersion spectrum for left and right surfaces, respectively, considering a number of discrete $k$-path segments as drawn inside a red circle in Fig.~\ref{fig:surface}(b).
Once we examine all $k$-lines connecting opposite points on the NR, the corresponding surface states together form the topologically nontrivial drumhead-like surface spectrum on both sides of the (001) slab.  
Next, for the $\kappa=-1$ AFM state in Fig.~\ref{fig:surface}(e), the topological charges of the  WPs are identified along the $\Gamma K M$ direction. 
These WPs are further projected to different surface momenta paths as shown in Fig.~\ref{fig:surface}(f), leading to surface energy spectrum with Fermi arcs as shown in Fig.~\ref{fig:surface}(g) and (h) for the left and the right surfaces, respectively.
Therefore, the projected surface band spectrum around the high-symmetry $K$-point confirms the existence of topologically nontrivial NRSM and WSM phases in the bulk Mn$_3$Sn associated with the $\kappa=+1$ and $\kappa=-1$ AFM configurations, respectively. Furthermore, we can conclude that by switching $\kappa$ from `$+1$' to `$-1$', the NR gets gapped out and evolves into a pair of WPs lying along the minor axis of the ellipse. 
The finding of such WPs is also consistent with earlier noncollinear AFM ground state \cite{Yang_2017}.

\begin{figure*}[t]
    \centering
     \includegraphics[width=7.0 in]{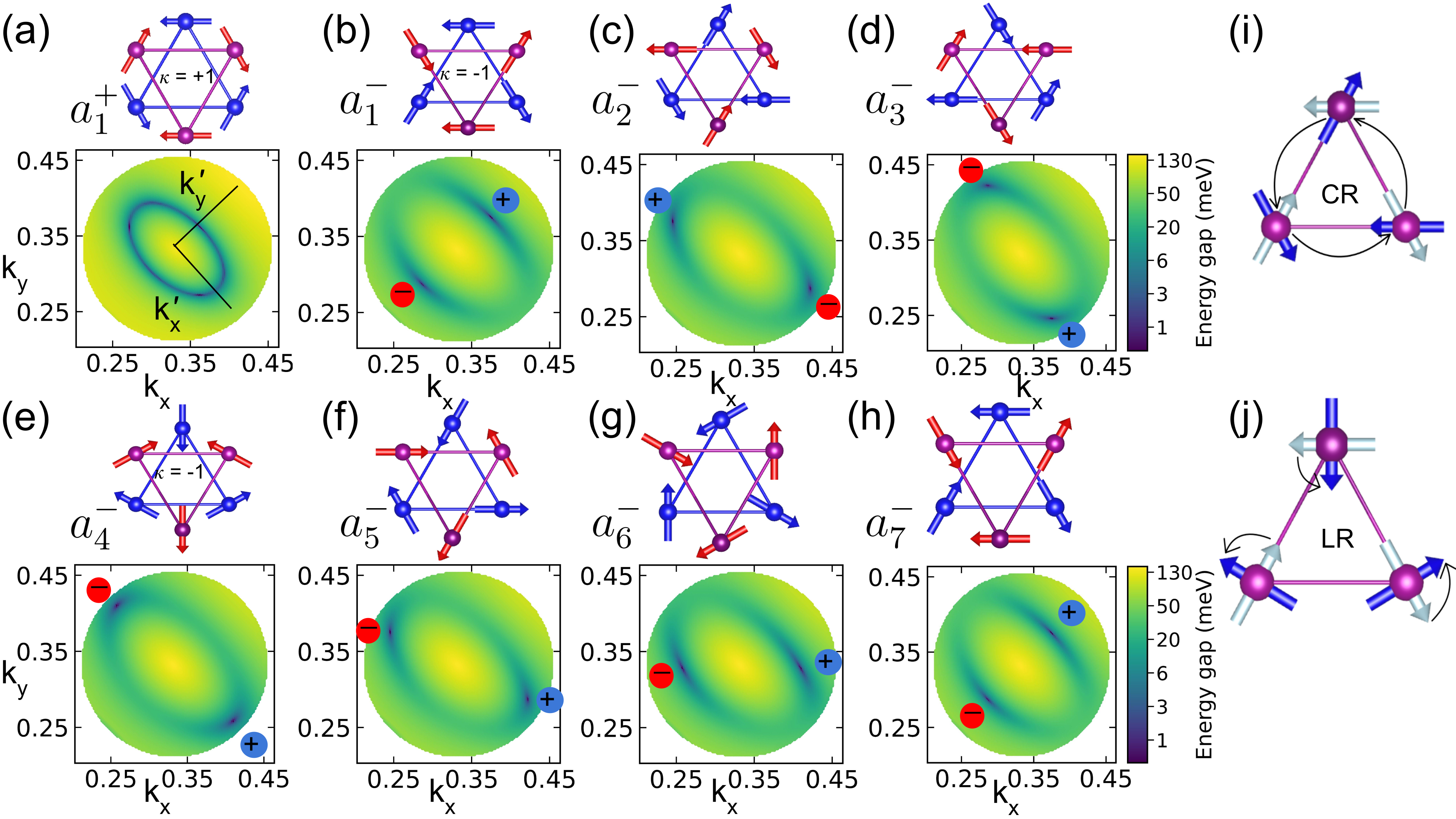}
     \caption{\textbf{Rotation of Weyl points and the associated tuning of anomalous Hall response.} Various topological semimetal phases, nodal-ring semimetal (NRSM) and Weyl semimetal (WSM) states, are calculated for different 120$^o$ noncollinear AFM states obtained by two types of collective rotations of \{\textbf{S}$_i$\}; crystalline rotation (CR) and local rotation (LR). {\bf (a)} Vector chirality $\kappa = +1$ state considering $a_1^+$ AFM order clearly shows NRSM state. {\bf (b)} $a_1^-$, {\bf (c)} $a_2^-$ and {\bf (d)} $a_3^-$ configurations all have vector chirality $\kappa=-1$, obtained by CR. Another set of $\kappa=-1$ states, {\bf (e)} $a_4^-$, {\bf (f)} $a_5^-$, {\bf (g)} $a_6^-$, {\bf (h)} $a_7^-$, are connected by clockwise LR of about $30^{o}$. Here, CR and LR both create Weyl-points at different locations of the ellipse. The locations and nontrivial topological charges of Weyl-points are calculated using the code WannierTools \cite{2018CoPhC.224..405W} based on the Wannier tight-binding model constructed using the Wannier90 \cite{MOSTOFI20142309}. {\bf (i)} and  {\bf (j)} are Schematic representations of CR and LR. } 
     \label{fig:WP}
\end{figure*}

\subsection{Stability of NR state, controlled generation (rotation) of WPs and the associated AHC response}

Here now, we explore interplay between the noncollinear AFM order dynamics and the associated TSM phases.
Keeping $\kappa$ unaltered, two additional magnetic textures are derived from the AFM configurations in Fig.~\ref{fig:mn3sn_a}(b) and (c) via collective co-rotations (same sense of rotation, here counterclockwise) of spins \{$\textbf{S}_i$\} in the unitcell by 90$^o$ around the local $z$-axis.
We find the NR is completely destroyed via fully opening up the gap for \textit{3-in-3-out} state ($\kappa=+1$) as indicated earlier in Fig.~\ref{fig:mn3sn_a}(i) and clearly visible in the Supplementary Figs.~3(e)-(h) with a discussion in Supplementary Note 2.
It is important to mention that the NRSM state survives only for a very special 120$^o$ noncollinear AFM arrangement.
Therefore, the stability of such NR state is inferred from subtle features in the noncollinear AFM configurations carrying $\kappa=+1$. 
On the contrary, the WPs in the other magnetic texture with $\kappa=-1$ are rotated by 90$^o$, which are now located at the major axis of the ellipse perpendicular to the $\Gamma K M$ line (see Supplementary Figs.~3(a)-(d)).
Note, the rotation of WPs has been proposed experimentally by electrical probing of the AHC in this material \cite{Tsai_2020}.

Furthermore, various noncollinear AFM textures keeping $\kappa$ constant to $-1$ are derived from collective co-rotations of \{$\textbf{S}_i$\}, either by 120$^o$ rotation with respect to the crystallographic $c$-axis (CR) as shown in Fig.~\ref{fig:WP}(i) or by continuous rotations with respect to the local $z$-axis (LR) as described in Fig.~\ref{fig:WP}(j). The calculated energy differences for all configurations in Fig.~\ref{fig:WP} with respect to $a_1^{+}$ are summarized in Table.~\ref{tab:my_label}. Each configuration with $\kappa=-1$ is now defined with an uniform LR angle $\phi$ which is zero for $a_4^{-}$ configuration and $a_5^{-}$ is connected with $a_4^{-}$ by 30$^o$ uniform LR. It is clearly evident that for configurations with odd multiplicative of rotation angle 30$^o$, $e.g.$ $a_1^{-}$/$a_7^{-}$, $a_2^{-}$, $a_3^{-}$, $a_5^{-}$ in Fig.~\ref{fig:WP}, the total energy is about 0.02 $meV/f.u.$ lower than that for the even multiplicative configurations, $e.g$ $a_4^{-}$, $a_6^{-}$ in Fig.~\ref{fig:WP}. Therefore, we observe two energetically competitive spin textures with the same $\kappa$ value. Although the energy difference is very small in our calculations, the Table.~\ref{tab:my_label} clearly indicates there are six energetically degenerate magnetic ground state as found within extended Heisenberg model~\cite{Sci.Adv.}. So, the presence of SOC in our calculations not only breaks the vector chiral symmetry (see the Supplementary Note 1 for more details), it also determines the degeneracy of the magnetic ground state.
The magnetic textures in Fig.~\ref{fig:WP}(b)-(h) confirm the existence of a pair of WPs with opposite topological charges, as shown in the bottom panel for each AFM state. Note that, all pairs of WPs are found lying on the perimeter of the elliptic NR. 
This, therefore, can be viewed as if the NRSM state is completely gapped out except a pair of points at different locations related to the magnetic textures.
However, to our surprise, the rotation of the WP pair goes in-phase with CR, (b)-(d), while out-of-phase with the LR, (e)-(h). In fact, the configurations $a_2^-$ and $a_5^-$ are connected through uniform LR of spins by 180$^o$, or equivalently by the time-reversal operation.
Hence, the manipulation of noncollinear AFM order allows one to access a plethora of quantum states which is expected to tailor various intrinsic electron transport properties.  

The NRSM to WSM phase transition by switching $\kappa$ as well as the rotation of WPs further leads us to explore how the controlled manipulation of chiral AFM order tailors the intrinsic AHC, $\sigma$. 
The Table~\ref{tab:my_label} shows the \textit{ab initio} calculated $\sigma_x$ and $\sigma_y$ components for all magnetic textures presented in the top panel of Fig.~\ref{fig:WP}(a)-(h).
The $\sigma$ turns out to be zero in case of NRSM state observed for $a_1^{+}$ spin configuration in Fig.~\ref{fig:WP}(a) as it generates zero Berry curvature around the NR \cite{article,PhysRevB.99.165117}.
In all noncollinear AFM configurations, $\sigma_z$ component is found to be zero.
Remarkably, a giant tunable $\sigma$ value is found once we switch to the WSM phase in the $\kappa=-1$ AFM states. 
Comparing $a_1^{-}$ and $a_4^-$ spin configurations in Fig.~\ref{fig:WP}(b) and (e), respectively and their corresponding nonzero values of $\sigma_x$ and $\sigma_y$, see Table.~\ref{tab:my_label}, we infer that the AHC is highly tunable depending on the noncollinear spin arrangements of AFMs. 
Furthermore, the sign difference in $\sigma_x$ and $\sigma_y$ values is clearly associated with the sense of rotation connecting $a_1^{-}$ and $a_4^-$, $e.g.$, a 90$^o$ counterclockwise LR (see also Supplementary Fig.~4 and Supplementary Note 2).
Likewise, the overall sign change in $\sigma$ is observed when one moves from AFM texture $a_2^-$ to $a_5^-$ by flipping all spins and each WP takes opposite chiral charge, see Fig.~\ref{fig:WP}(c) and (f).  
Therefore, by controlling noncollinear magnetic order in Mn$_3$Sn, the TSM phases can be altered between two semimetals, one carries NR state while the other one carries rotating WPs.
This, in turn, opens the possibility of switching $\sigma$ from zero to a giant value as well as tailoring its components in a single material.

\begin{table}[]
\begin{ruledtabular}
  \begin{tabular}{||c c c c||}
    \hline
      Spin 
    & Energy/f.u.
    & $\sigma_x$
    & \textbf{$\sigma_y$} \\ [0.5ex]
     configuration
    & (meV)
    & $(S\cdot cm^{-1})$
    & $(S\cdot cm^{-1})$ \\ 
    \hline \hline
    $a_1^{+}$
    & ~~0 
    & ~~0
    & ~~0 \\
    \hline
    $a_1^{-} ~~(\phi = 90^{o})$
    & -4.472
    & -226
    &~~0 \\
    \hline
    ~$a_2^{-}~~(\phi= 210^{o})$
    & -4.472
    & ~116
    & -204 \\
    \hline
    ~$a_3^{-}~~ (\phi= 330^{o})$
    & -4.472
    & ~116
    & ~204 \\
    \hline
    $a_4^{-}~~~ (\phi= 0^{o})$
    & -4.451
    & ~~0
    & ~223 \\
    \hline
    $a_5^{-}~~ (\phi = 30^{o})$
    & -4.472
    & -116
    & ~204 \\
    \hline
    $a_6^{-} ~~ (\phi = 60^{o})$
    & -4.451
    & -200
    & -104 \\
    \hline
    $a_{7}^{-}~~ (\phi = 90^{o})$
    & -4.472
    & -226
    & ~~0 \\
    \hline
\end{tabular}
\end{ruledtabular}
\caption{\textbf{Calculated anomalous Hall conductivity for different noncollinear antiferromagnetic phases.} The stability energies of vector chirality $\kappa=-1$ noncollinear antiferromagnetic states with respect to $\kappa=+1$ one and the corresponding intrinsic anomalous Hall conductivity, $\sigma_{x(y)}$. The angle $\phi$ in parenthesis is defined with respect to the $a_4^{-}$ antiferromagnetic configuration, representing uniform spin rotation measured in integer multiple of 30$^o$.}
\label{tab:my_label}
\end{table}
%

\section{Discussion}
In Mn$_3$Sn, the essential topological footprints near the $E_f$ are captured by constructing a low-energy $\textbf{k.p}$ effective model Hamiltonian of two bands around the high-symmetric $K$ points in the BZ.
Here, in the construction, we take into account all relevant symmetries in the noncollinear AFMs and the detailed construction of the theory is given in the Supplementary Note 3.
The Hamiltonian for the existence of a NR solution takes the form
\begin{equation}
    \mathcal{H}_\textrm{NR} = a_1 k_z^{2} k_x \sigma_x + (m_0 - m_1 k_x^{2} - m_2 k_y^{2} - m_3 k_z^{2}) \sigma_z,
    \label{Eq:NR}
\end{equation}
where \{$\sigma_i$\} are the Pauli matrices and the momentum, $\mathbf{k}$ = ($\textbf{k}_{||}, k_z$), is measured relative to the $K$ point.
$a$ and $m$ are the free parameters in the model. 
The eigenvalue solutions, $\mathcal{E}_{\pm \mathbf{k}}$, are degenerate in the $k_z = 0$ plane, forming an elliptical NR state satisfying $m_1 k_x^2 + m_2 k_y^2 = m_0$.
The constructed Hamiltonian is invariant under important symmetries, $\mathcal{C}_{3z}$, $\mathcal{M}_x$, $\mathcal{M}_y\mathcal{T}$ and $\mathcal{M}_z\mathcal{T}$, observed in the particular $\kappa = +1$ AFM state, $a_1^+$ in Fig.~\ref{fig:WP}(a) (also see Fig.~\ref{fig:mn3sn_a}(b)).
As stated earlier, a simultaneous local rotation of \{$\textbf{S}_i$\} in the unit cell keeps $\kappa$ value unchanged along with the $\mathcal{C}_{3z}$ rotation symmetry intact, however, can change the other symmetries.
If we replace the $a_1 k_z^{2} k_x \sigma_x$ in Eq.~\ref{Eq:NR} with $a_1 \mathbf{k}^2_{||} k_y \sigma_x$, the new $\textbf{k.p}$ Hamiltonian on the hand preserves symmetries, $\mathcal{M}_x T$, $\mathcal{M}_y$, $\mathcal{M}_z T$  and $\mathcal{C}_{3z}$ which are protected in another $\kappa=+1$ AFM state, \textit{3-in-3-out} configuration.
In that case, the NR solution disappears as the bands are completely gapped out for all $\textbf{k}_{||}$ points.
The \textit{ab initio} band structure is in good agreement with the model solution and the fully gapped state  is clearly visible in the 3D energy band dispersion, see Supplementary Fig.~3(h). 
Therefore, the existence of NR solution on the $k_z=0$ plane is found for the coexistence of $\mathcal{C}_{3z}$, $\mathcal{M}_x$, $\mathcal{M}_y\mathcal{T}$  and $\mathcal{M}_z\mathcal{T}$ symmetries, particularly preserved in $a_1^+$ magnetic texture.
Above combination of symmetries is very unique as moving to $\kappa=-1$ AFM state, $a_1^-$ in Fig.~\ref{fig:WP}(b) (constructed via $120^o~staggered$ rotation), $\mathcal{M}_x$, $\mathcal{M}_y\mathcal{T}$ and $\mathcal{M}_z\mathcal{T}$ symmetries remain conserved while the $\mathcal{C}_{3z}$ symmetry is broken. 
The effective low energy Hamiltonian in that case takes the form
\begin{equation}
    \mathcal{H}_\textrm{WP} = v_x k_x \sigma_x +(m_0 - m_1 k_x^{2} -m_2 k_y^2 - m_3 k_z^{2}) \sigma_z,
    \label{Eq:WN}
\end{equation}
where the $v_x k_x \sigma_x$ is the perturbing term. 
The energy dispersion solution for $k_z=0$ plane shows a pair of gapless points at $\mathbf{k}_{||}=(0,\pm \sqrt{m_0/m_2})$ along the $k_y$ axis which is the minor axis of the elliptic NR solution earlier.
Therefore, we infer that the perturbation leads to generate two crossing points on the minor axis of the ellipse, consistent with the \textit{ab initio} band structure.
On the other hand, by replacing $v_x k_x \sigma_x$ to $v_y k_y\sigma_x$ in Eq.~\ref{Eq:WN}, we find a pair of gapless points at $\textbf{k}_{||}=(\pm \sqrt{m_0/m_1},0)$ along the $k_x$ axis.
This Hamiltonian is invariant under $\mathcal{M}_x T$, $\mathcal{M}_y$ symmetries which are conserved for $a_4^-$ AFM state in Fig.~\ref{fig:WP}(e).
This is hence consistent with the location of the \textit{ab initio} calculated WPs, lying along the major axis of the ellipse. The rotation of WPs related to the collective spin rotations, therefore, has led to choose a perturbation of the form ($\mathbf{v\cdot k}_{||})\sigma_x$ in Eq.~\ref{Eq:WN}.
Most importantly, a pair of gapless points is always found as the solution of the $\mathbf{k}\cdot\mathbf{p}$ Hamiltonian in the $k_x$-$k_y$ plane. 
Depending on the vector $\mathbf{v}$ in the Hamiltonian, a pair of band touching points in the solution change their positions on an elliptic ring. 
This is consistent with the rotation of the WPs around $K$ point in the band structure of various magnetic textures with $\kappa = -1$. 
The vector $\mathbf{v}$ here can be closely associated with the magnetic octupole moments in Mn$_3$Sn which is believed to break the time reversal symmetry \cite{Tsai_2020, PhysRevB.95.094406}.  

In our results, we have established the intimate connection between vector chirality driven topological phase transition and AHE in antiferromagnets. 
Importantly, the noncollinear AFM spin texture has a significant impact on its band structure, Berry curvature and hence, on the intrinsic AHE.  
The NR state in case of $a_1^+$ is found to exhibit net zero Berry curvature in its vicinity and thus does not generate any AHE.
We also find that the fully gapped NR state found in \textit{3-in-3-out} AFM texture does not induce Berry curvature, which is responsible for the absence of AHE.
This therefore, is in contrast with the ferromagnetic topological semimetal Co$_2$MnAl \cite{article} that generate large AHE only by gapping out NRs in the presence of net magnetization in the system.
On the other hand, the elliptic NR has evolved into a pair of WPs when the $a_1^+$ texture turns into $a_1^-$ texture by $120^o$ $staggered$ rotation.
The WPs contribute in generating large Berry curvature, which is responsible for the observed giant AHC. 
The absence of intrinsic AHE may arise either from nontrivial topological NR phase or from the gapped phase, depending on the underlying noncollinear AFM texture.
Nonetheless, the switching of AHC from zero to a giant value may indirectly justify our claim of topological phase transition from the NRSM to the WSM phase. 
Moreover, the smooth variation of $\sigma_x$ or $\sigma_y$ component (including switching between  positive value to negative value and vice-versa) is consistent with the rotation of the WPs on an elliptic ring, compare the results in Fig.~\ref{fig:WP} and Table~\ref{tab:my_label}.
The important point to be noticed is that the giant $\sigma_x$ ($\sigma_y$) value, particularly in case of $a_1^{-}$ ($a_4^-$) AFM state, arises due to the odd nature of the Berry curvature $\Omega_y$ ($\Omega_x$) under the symmetry operation $\mathcal{M}_yT$ ($\mathcal{M}_xT$).
Therefore, the odd Berry curvature does not contribute in the AHE.
In case of intermediate $120^o$ AFM configurations ($e.g.$ $a_5^-$ and $a_6^-$ connected by LR), the relevant symmetries are broken which then lead to nonzero values of $\sigma_x$ and $\sigma_y$ both.

In summary, based on the detailed electronic structure calculations and low energy $\mathbf{k}\cdot\mathbf{p}$ effective theory, we systematically reveal that multiple nontrivial TSM phases can be realized in $120^o$ noncollinear AFM, Mn$_3$Sn, by controlled alternation of the spin configurations. 
The analyses demonstrate that the NR and the gapless nodes in the form of WPs are strongly rely on the chiral orders characterized by the vector chirality, $\kappa=+1$ and $-1$, respectively.
This work suggests that a $staggered$ rotation involving two spins in the kagome triangle is an unconventional route to topological phase transition from an elliptical NR to a pair of WPs through the switching of $\kappa$ from `$+1$' to `$-1$'.
It is worth to mention that the $staggered$ torque with sufficient strength required for switching $\kappa$ is difficult to generate at low temperature. However, at elevated temperatures when multiple magnetic phases may coexist \cite{apl2018}, the required torque can be reduced significantly. Notably, such types of $staggered$ rotations, albeit small in magnitude, have been recently realized in this class of materials via an external magnetic field  \cite{miwa} and uniaxial strain \cite{pizo}.
Later, depending on the nature of rotation (CR or LR) of spins in the $\kappa=-1$ AFM texture, the WPs are found to rotate as if the topological charges can be created on preferred locations on the ring.
The symmetry adapted $\mathbf{k}\cdot\mathbf{p}$ theory moreover captures our findings in good agreement and, hence, brings an unconventional way of generating a pair of WPs from a ring in noncollinear AFMs.
We further find a remarkable switching of AHC from zero to a giant value associated with the TSM phase transition in the system carrying nonzero vector chirality. 
Finally, it shows that the components of AHC, $\sigma_x$ and $\sigma_y$, can be tuned smoothly depending on the location of the WPs on an elliptic ring. These topological features in the band structures are also observed in the sister compounds in the hexagonal Mn$_3$X family (see the discussion in Supplementary Note 4 and Supplementary Figs.~5 and 6 for Mn$_3$Ge and Mn$_3$Ga, respectively).
We thus believe that the TSM phase dynamics in hexagonal $120^o$ noncollinear AFM will offer a new avenue to develop concepts on $staggered$ torques for the manipulation of vector chiral order in AFMs and it possibly adds an alternative component to \textit{antiferromagnetic spintronics}.

\section{Methods}
\subsection*{Ab initio electronic structure calculations}

Density Functional Theory (DFT) calculations are carried out with two different approaches: the full-potential linearized augmented plane wave (FLAPW) method as implemented in the J\"{u}lich DFT code FLEUR  \cite{fleur}, and the plane-wave projected augmented wave (PAW) method as implemented in Vienna ab initio Simulation Package (VASP) \cite{vasp,paw2}.
We have carefully checked the consistency of our calculations in the above mentioned approaches in terms of density of states, band structures and stability of magnetic states. 
The total energy calculations for different noncollinear antiferromagnetic (AFM) structures with and without spin-orbit coupling (SOC), are carried out in the plane wave basis with projector-augmented wave (PAW) potentials. 
A plane-wave cutoff of 500 eV and $\Gamma$-centered $k$-mesh of $8\times8\times9$ are found to provide a good convergence of the total energies. 
On the other hand, for the self-consistent calculations in FLEUR, we consider a plane-wave cutoff of $k_{max}= 4.2$\,a.u.$^{-1}$ for expanding the LAPW basis functions where the charge densities are converged using a Monkhorst-Pack \cite{monk} $k$-mesh of $8 \times 8\times 9$ in the whole Brillouin zone (BZ). 
The muffin-tin radii for Mn and Sn are set to 2.57\, a.u. and 2.64\,a.u., respectively.
We use the Vosko-Wilk-Nusair (VWN) \cite{VWN} exchange-correlation functional within the local density approximation (LDA) for the self-consistent calculations.
The plane wave cutoff for the potential ($g_{max}$) and exchange-correlation potential ($g_{max,xc}$) are chosen to be 14.0 and 12.0\,a.u.$^{-1}$, respectively. These choices of the numerical parameters are found to provide good convergence of the total energy.
Our calculations include the effect of SOC self-consistently.
The total energies calculated in both FLEUR and VASP are consistent and comparable.
The calculated total energy difference in the presence of SOC between $\kappa=+1$ and $-1$ noncollinear AFM states is 3.8 meV/f.u. within FLEUR whereas it is 4.5 meV/f.u. in VASP calculation.
To construct the three-dimensional (3D) energy band dispersion of each noncollinear AFM state using {\it ab initio} method, we take a very dense circular $k$-mesh around the high symmetry point K. The electronic band structures with SOC are further parameterized with maximally-localized Wannier functions (MLWFs) \cite{MOSTOFI20142309} within all electron full potential methods of LAPW as implemented in FLEUR \cite{fleur}.
Atomic orbital-like MLWFs of Mn-$d$, Sn-$p$ states are considered to construct the tight-binding (TB) Hamiltonian, which reproduces the spectrum of the system accurately within a large energy window ($\approx$ 7.5\,eV) around the Fermi energy.
From the constructed TB Hamiltonian based on MLWFs as implemented in the WannierTools software \cite{2018CoPhC.224..405W}, the surface spectrum for (001) surfaces is calculated using Green’s function iterative approach. 
The position of Weyl-nodes and their topological charges are also calculated using the same tool.
The Fermi arcs connecting two Weyl-points are clearly identified in the surface band spectrum.

Then to compute the anomalous Hall conductivity, we evaluate the intrinsic Berry curvature contribution employing the Wannier interpolation technique \cite{wan1} as implemented in the FLEUR code \cite{Freimuth-2008}. 
The Berry curvatures are computed from a well constructed TB-model based on the MLWFs \cite{wan2}. 
The linear response Kubo formula approach \cite{Yao-2004} has been employed as follows:
\begin{eqnarray}
\Omega_{n}(\mathbf{k}) = -\hbar^{2} \sum_{n \neq m} \frac{\operatorname{2 Im} \langle u_{n\mathbf{k}}|\hat{v}_{i}|u_{m\mathbf{k}}\rangle \langle u_{n\mathbf{k}}|\hat{v}_{j}|u_{m\mathbf{k}}\rangle}{(\epsilon_{n\mathbf{k}}-\epsilon_{m\mathbf{k}})^{2}},
\end{eqnarray}
where $\Omega_{n}(\mathbf{k})$ is the Berry curvature of band $n$, $\hat{v}_{i} =\frac{1}{\hbar}{\partial \hat{H}(\mathbf{k})}/{\partial k_{i}} $ is the velocity operator with $i\in \{x,y\}$, $u_{n\mathbf{k}}$ and $\epsilon_{n\mathbf{k}}$ are the eigenstates and eigenvalues of the Hamiltonian $\hat{H}(\mathbf{k})$, respectively.
Subsequently, we calculate the anomalous Hall conductivity (AHC) as given by: 
\begin{equation}
\begin{aligned}
\sigma^{A}_{H}=& -\hbar e^{2} \sum_{n} \int_{BZ} \frac{d{\bf k}}{\left(2 \pi\right)^3}f_{n}(\mathbf{k})\Omega_{n}(\mathbf{k}),
\end{aligned}
\end{equation}
To compute $\sigma_H^A$, we use a very dense $k$-mesh of $300 \times 300 \times 300$, and such a dense mesh is found to give well-converged values of the AHC. 

\section*{Data Availability}
The data of this study are available from the corresponding author upon reasonable request via email: aknandy@niser.ac.in.

\section*{Code Availability}
The DFT calculations are performed with public codes. Their input files are available upon reasonable request.

\section*{Acknowledgment}
A.K.N, S.P. and K. Saha acknowledge the support from the Department of Atomic Energy, Government of India. A.K.N. and S.P. acknowledge the computational resources, Kalinga cluster, at National Institute of Science Education and Research, Bhubaneswar, India. A.K.N. thanks Prof. P. M. Oppeneer for the Swedish National Infrastructure for Computing (SNIC) facility. A.K.N. thanks Prof. P. Mahadevan for critical reading of the manuscript and stimulating discussions. A.K.N. and S.P. acknowledge Dr. Ajaya K. Nayak, Charanpreet Singh, Dr. Hirak K. Chandra, Sandip Bera, Sayan Banik, Arghya Mukherjee for fruitful discussion. K. Samanta acknowledge the computing resources granted by JARA-HPC from RWTH Aachen University and Forschungszentrum J\"ulich, Germany.

\section*{Author Contributions}
A.K.N. has conceptualized and supervised the work. The ab initio simulations are performed by S.P., K.Samanta and A.K.N. The \textbf{k.p} theory is constructed by S.P., K.Saha and A.K.N. The paper was written by A.K.N with inputs from S. P., K.Samanta and K.Saha. All authors have contributed to the discussions and analyses of the data and approved the final version.

\section*{Competing Interests}
The authors declare no competing financial interests.
\section*{References}
\providecommand{\noopsort}[1]{}\providecommand{\singleletter}[1]{#1}%


\begin{thebibliography}{10}

\bibitem{RevModPhys.90.015001}
 Armitage, N. P., Mele, E. J. \& Vishwanath, A.
\newblock Weyl and dirac semimetals in three-dimensional solids.
\newblock {\em Rev. Mod. Phys.} \textbf{90}, 015001 (2018).

\bibitem{doi:10.1146/annurev-matsci-070218-010049}
Gao, Heng., Venderbos, J.~W., Kim, Y. \& Rappe, A.~M.
\newblock Topological semimetals from first principles.
\newblock {\em Annu. Rev. Mater. Res.} \textbf{49}, 153-183 (2019).

\bibitem{Hasan_2021}
Hasan, M.~Z., Chang, G., Belopolski, I., Bian, G., Xu,  S.-Y. \&
Yin, J.-X.
\newblock Weyl, dirac and high-fold chiral~fermions in topological quantum matter.
\newblock {\em Nat. Rev. Mater.} \textbf{6}, 784-803 (2021).

\bibitem{Yang_2016}
Yang, S.
\newblock Dirac and weyl materials: Fundamental aspects and some spintronics
  applications.
\newblock {\em {SPIN}}, \textbf{06}, 1640003, (2016).

\bibitem{PhysRevLett.108.140405}
Young, S.~M., Zaheer, S., Teo, J.~C.~Y., Kane, C.~L., Mele, E.~J., \&  Rappe, A.~M.
\newblock Dirac semimetal in three dimensions.
\newblock {\em Phys. Rev. Lett.} \textbf{108}, 140405 (2012).

\bibitem{PhysRevB.83.205101}
Wan, X. G., Turner, A. M., Viswanath, A. \& Savrasov, S. Y.
\newblock Topological semimetal and fermi-arc surface states in the electronic structure of pyrochlore iridates.
\newblock {\em Phys. Rev. B} \textbf{83}, 205101 (2011).

\bibitem{PhysRevB.99.075116}
Nandy, S., Saha, K., Taraphder, A. \& Tewari, S.
\newblock Mirror anomaly and anomalous hall effect in type-1 dirac semimetals.
\newblock {\em Phys. Rev. B} \textbf{99}, 075116 (2019).

\bibitem{PhysRevB.92.045108}
Weng, H. et al.
\newblock Topological node-line semimetal in three-dimensional graphene networks.
\newblock {\em Phys. Rev. B} \textbf{92}, 045108 (2015).

\bibitem{Fang_2016}
Fang, C., Weng,H., ~Dai, X. \& Fang, Z.
\newblock Topological nodal line semimetals.
\newblock {\em Chinese Phys. B} \textbf{25}, 117106 (2016).

\bibitem{PhysRevLett.122.077203}
 Shao, D.-f., Gurung, G., Zhang, S.-h \& Tsymbal, E. Y.
\newblock Dirac nodal line metal for topological antiferromagnetic spintronics.
\newblock {\em Phys. Rev. Lett.} \textbf{122}, 077203 (2019).

\bibitem{Belopolski_2019}
Belopolski, I. et al.
\newblock Discovery of topological weyl fermion lines and drumhead surface states in a room temperature magnet.
\newblock {\em Science} \textbf{365}, 1278-1281 (2019).

\bibitem{PhysRevB.104.235136}
Li, J., Wang, H. \&  Pan, H.
\newblock Tunable topological phase transition from nodal-line semimetal to
  weyl semimetal by breaking symmetry.
\newblock {\em Phys. Rev. B} \textbf{104}, 235136 (2021).

\bibitem{PhysRevB.97.075120}
T\"urker, O.b.u. \& Moroz,  S.
\newblock Weyl nodal surfaces.
\newblock {\em Phys. Rev. B} \textbf{97}, 075120 (2018).

\bibitem{D0TC01978J}
Yang, T. \& Zhang, X.
\newblock Nearly flat nodal surface states in pseudo-one-dimensional molybdenum monochalcogenides X(MoS)$_3$ (X=K,Rb and Cs).
\newblock {\em J. Mater. Chem. C} \textbf{8}, 9046-9054 (2020).

\bibitem{PhysRevB.97.235150}
Zhang, X. et al.
\newblock Nodal loop and nodal surface states in the ${\mathrm{Ti}}_{3}\mathrm{Al}$ family of materials.
\newblock {\em Phys. Rev. B} \textbf{97}, 235150 (2018).

\bibitem{huang2015weyl}
Huang, SM. et al.
\newblock A weyl fermion semimetal with surface fermi arcs in the transition
  metal monopnictide TaAs class.
\newblock {\em Nat. Commun.} \textbf{6}, 1-6 (2015).

\bibitem{Yang_2017}
Yang, H., Sun, Y., Zhang, Y., Shi, W.-J., Parkin, S.~S.~P.
\& Yan, B.
\newblock Topological weyl semimetals in the chiral antiferromagnetic materials Mn$_3$Ge and Mn$_3$Sn.
\newblock {\em New Journal of Physics} \textbf{19}, 015008 (2017).

\bibitem{Kuroda_2017}
Kuroda, K et al.
\newblock Evidence for magnetic weyl fermions in a correlated metal.
\newblock {\em Nat. Mater.} \textbf{16}, 1090-1095 (2017).

\bibitem{Yan_2017}
Yan, B. \& Felser, C.
\newblock Topological materials: Weyl semimetals.
\newblock {\em Annu. Rev. Condens. Matter Phys.} \textbf{8}, 337-354 (2017).

\bibitem{Liu_2014}
Liu, Z.~K. et al.
\newblock Discovery of a three-dimensional topological dirac semimetal, Na$_3$Bi.
\newblock {\em Science}, \textbf{343}, 864-867 (2014).

\bibitem{doi:10.1021/jacs.8b09900}
Jing, Yu. \&  Heine, T.
\newblock Two-dimensional kagome lattices made of hetero triangulenes are dirac semimetals or single-band semiconductors.
\newblock {\em Journal of the American Chemical Society} \textbf{141}, 743-747 (2019).

\bibitem{C3CP53257G}
Liu, Z., Wang, J. \& and Li., J.
\newblock Dirac cones in two-dimensional systems: from hexagonal to square lattices.
\newblock {\em Phys. Chem. Chem. Phys.} \textbf{15}, 18855-18862 (2013).

\bibitem{hosen2018distinct}
 Hosen, M.M. et al.
\newblock Distinct multiple fermionic states in a single topological metal.
\newblock {\em Nat. Commun.} \textbf{9}, {1-8} (2018).

\bibitem{2019NatRP...1..126T}
Tokura, Y., Yasuda, K. \& Tsukazaki, A.
\newblock {Magnetic topological insulators}.
\newblock {\em Nat. Rev. Phys.} \textbf{1}, 126-143 (2019).

\bibitem{naturematerial}
He, Q., Hughes, T., Armitage, N., Tokura, Y. \& Wang, K. 
\newblock Topological spintronics and magnetoelectronics.
\newblock {\em Nat. Mater.} \textbf{21}, 15-23 (2022).

\bibitem{PhysRevLett.115.036805}
Wang, J., Lian, B. \& Zhang, S-C.
\newblock Electrically tunable magnetism in magnetic topological insulators.
\newblock {\em Phys. Rev. Lett.} \textbf{115}, 036805 (2015).

\bibitem{zhang2014electrically}
Zhang, Z. et al.
\newblock Electrically tuned magnetic order and magnetoresistance in a topological insulator.
\newblock {\em Nat. Commun.}, \textbf{5}, 1-7 (2014).

\bibitem{doi:10.1073/pnas.1713458114}
Yeats, A. et al.
\newblock Local optical control of ferromagnetism and chemical potential in a
topological insulator.
\newblock {\em PNAS}
\textbf{114}, 10379-10383 (2017).

\bibitem{article}
Li, Peigang. et al.
\newblock Giant room temperature anomalous hall effect and tunable topology in a ferromagnetic topological semimetal Co$_2$MnAl.
\newblock {\em Nat. Commun.} \textbf{11}, 3476 (2020).

\bibitem{PhysRevX.9.041040}
Chen, Y.~J. et al.
\newblock Topological electronic structure and its temperature evolution in antiferromagnetic topological insulator ${\mathrm{MnBi}}_{2}{\mathrm{Te}}_{4}$.
\newblock {\em Phys. Rev. X} \textbf{9}, 041040 (2019).

\bibitem{Otrokov_2019}
Otrokov, M. et al.
\newblock Prediction and observation of an antiferromagnetic topological insulator.
\newblock {\em Nature} \textbf{576}, 416-422, 12 (2019).

\bibitem{wangC}
Wang, C., Gao, Y. \& Xiao, D.
\newblock Intrinsic nonlinear Hall effect in antiferromagnetic tetragonal CuMnAs.
\newblock {\em Phys. Rev. Lett.} \textbf{127}, 277201 (2021).

\bibitem{liuH}
Liu, H. et al.
\newblock Intrinsic second-order anomalous Hall effect and its application in compensated antiferromagnets.
\newblock {\em Phys. Rev. Lett.} \textbf{127}, 277202 (2021).

\bibitem{Wadley_2016}
Wadley, P. et al.
\newblock Electrical switching of an antiferromagnet.
\newblock {\em Science} \textbf{351}, 587-590 (2016).

\bibitem{Zelezny}
Železný, J. et al.
\newblock Relativistic Néel-order fields induced by electrical current in antiferromagnets.
\newblock {\em Phys. Rev. Lett.} \textbf{113}, 157201 (2014).

\bibitem{Nature_anti}
Smejkal, L., Mokrousov, Y., Yan, B. \& MacDonald, A~H.
\newblock Prediction and observation of an antiferromagnetic topological insulator.
\newblock {\em Nature}, \textbf{14}, 242–251 (2018).

\bibitem{prl118}
Šmejkal, L., Železný, J., Sinova, J. \& Jungwirth, T.
\newblock Electric control of Dirac quasiparticles by spin-orbit torque in an antiferromagnet.
\newblock {\em Phys. Rev. Lett.} \textbf{118}, 106402 (2017).

\bibitem{diang-Fu2019}
Shao, D., Gurung, G., Zhang, S \& Tsymbal, E. Y.
\newblock Dirac nodal line metal for topological antiferromagnetic spintronics.
\newblock {\em Phys. Rev. Lett.} \textbf{122}, 077203 (2019).

\bibitem{salemi2019orbitally}
Salemi, L., Berritta,  M., Nandy, A.~K. \& Oppeneer P.~M.
\newblock Orbitally dominated rashba-edelstein effect in noncentrosymmetric antiferromagnets.
\newblock {\em Nat. Commun.}, \textbf{10}, 1-10 (2019).

\bibitem{PhysRevLett.112.017205}
Chen, H., Niu, Q. \& MacDonald, A.H.
\newblock Anomalous hall effect arising from noncollinear antiferromagnetism.
\newblock {\em Phys. Rev. Lett.} \textbf{112}, 017205 (2014).

\bibitem{PhysRevB.92.144426}
Feng, W., Guo, G.-Y., Zhou, J., Yao, Y. \& Niu, Q.
\newblock Large magneto-optical kerr effect in noncollinear antiferromagnets
  ${\mathrm{Mn}}_{3}X\phantom{\rule{0.28em}{0ex}}(X=\mathrm{Rh},\phantom{\rule{0.28em}{0ex}}\mathrm{Ir},\phantom{\rule{0.28em}{0ex}}\mathrm{Pt})$.
\newblock {\em Phys. Rev. B} \textbf{92}, 144426  (2015).

\bibitem{K_bler_2014}
Kübler, J. \& Felser, C.
\newblock Non-collinear antiferromagnets and the anomalous hall effect.
\newblock {\em {EPL} (Europhysics Letters)} \textbf{108}, 67001 (2014).

\bibitem{nakatsuji2015large}
Nakatsuji, S., Kiyohara, N. \& Higo, T. 
\newblock Large anomalous hall effect in a non-collinear antiferromagnet at
  room temperature.
\newblock {\em Nature}, \textbf{527}, 212-215, (2015).

\bibitem{doi:10.1126/sciadv.1501870}
Nayak, A. K. et al.
\newblock Large anomalous hall effect driven by a nonvanishing berry curvature
  in the noncolinear antiferromagnet Mn$_3$Ge.
\newblock {\em Sci. Adv.}. \textbf{2}, e1501870 (2016).

\bibitem{doi:10.1063/1.5143959}
Wu, M., Isshiki, H., Chen, T., Higo, T., Nakatsuji, S. \&
Otani, Y.
\newblock Magneto-optical kerr effect in a non-collinear antiferromagnet Mn$_3$Ge.
\newblock {\em Appl. Phys. Lett.} \textbf{116}, 132408 (2020).

\bibitem{PhysRevB.95.075128}
Zhang, Y. et al.
\newblock Strong anisotropic anomalous hall effect and spin hall effect in the chiral antiferromagnetic compounds ${\mathrm{Mn}}_{3}X$ ($X=\mathrm{Ge}$, Sn,
  Ga, Ir, Rh, and Pt).
\newblock {\em Phys. Rev. B} \textbf{95}, 075128 (2017).

\bibitem{Ikhlas_2017}
Ikhlas, M. et al.
\newblock Large anomalous nernst effect at room temperature in a chiral antiferromagnet.
\newblock {\em Nat. Phys.} \textbf{13}, 1085-1090 (2017).

\bibitem{csingh}
Singh, C. et al.
\newblock Higher order exchange driven noncoplanar magnetic state and large anomalous Hall effects in electron doped kagome magnet Mn$_3$Sn
\newblock {\em arXiv. https://arxiv.org/abs/2211.12722} (2022).

\bibitem{PhysRevLett.76.4963}
Sandratskii, L. M. \& Kübler, C.
\newblock Role of Orbital Polarization in Weak Ferromagnetism.
\newblock {\em Phys. Rev. Lett.} \textbf{76}, 4963 (1996).


\bibitem{Tsai_2020}
Tsai, H. et al.
\newblock Electrical manipulation of a topological antiferromagnetic state.
\newblock {\em Nature} \textbf{580}, 608-613 (2020).

\bibitem{PhysRevB.99.104428}
Zhou, X. et al.
\newblock Spin-order dependent anomalous hall effect and magneto-optical effect in the noncollinear antiferromagnets ${\mathrm{Mn}}_{3}X\mathrm{N}$ with $X=\mathrm{Ga}$, Zn, Ag, or Ni.
\newblock {\em Phys. Rev. B} \textbf{99}, 104428 (2019).

\bibitem{Kawamura_2001}
Kawamura, H.
\newblock Spin- and chirality-orderings of frustrated magnets stacked-triangular anti-ferromagnets and spin glasses.
\newblock {\em Can. J. Phys.} \textbf{79}, 1447-1458 (2001).

\bibitem{Zhangprb}
Zhang, Y. et al.
\newblock Strong anisotropic anomalous Hall effect and spin Hall effect in the chiral antiferromagnetic compounds  Mn$_3$X (X=Ge, Sn, Ga, Ir, Rh, and Pt).
\newblock {\em Phys. Rev. B} \textbf{95}, 075128 (2017).

\bibitem{PhysRevB.95.094406}
Suzuki, M.-T., Koretsune, T., Ochi, M. \& Arita, R.
\newblock Cluster multipole theory for anomalous hall effect in antiferromagnets.
\newblock {\em Phys. Rev. B} \textbf{95}, 094406 (2017).

\bibitem{prb101}
Soh, J.-R. et al.
\newblock Ground-state magnetic structure of  Mn$_3$Ge.
\newblock {\em Phys. Rev. B} \textbf{101}, 140411(R) (2020).

\bibitem{fleur}
{www.flapw.de}.

\bibitem{vasp}
Kresse, G. \& Joubert,  D.
\newblock From ultrasoft pseudopotentials to the projector augmented-wave method.
\newblock {\em Phys. Rev. B} \textbf{59}, 1758-1775 (1999).

\bibitem{paw2}
 Bl\"ochl, P.~E.
\newblock Projector augmented-wave method.
\newblock {\em Phys. Rev. B} \textbf{50}, 17953-17979, (1994).

\bibitem{PhysRevResearch.2.043366}
Singh, C. et al.
\newblock Pressure controlled trimerization for switching of anomalous hall
  effect in triangular antiferromagnet ${\mathrm{Mn}}_{3}\mathrm{Sn}$.
\newblock {\em Phys. Rev. Research} \textbf{2}, 043366 (2020).

\bibitem{chen2021anomalous}
Chen, T. et al.
\newblock Anomalous transport due to weyl fermions in the chiral antiferromagnets Mn$_3$X, X= Sn, Ge.
\newblock {\em Nat. Commun.}, \textbf{12}, 1-14 (2021).

\bibitem{MOSTOFI20142309}
Mostofi, A. A. et al.
\newblock An updated version of wannier90: A tool for obtaining maximally-localised wannier functions.
\newblock {\em Comput. Phys. Commun}. \textbf{185}, 2309-2310 (2014).

\bibitem{2018CoPhC.224..405W}
Wu, Q., Zhang, S., Song, H.-F., Troyer, M. \& Soluyanov, A.~A. 
\newblock {WannierTools: An open-source software package for novel topological
  materials}.
\newblock {\em Comput. Phys. Commun.} \textbf{224}, 405-416 (2018).

\bibitem{Sci.Adv.}
Pal, B. et al.
\newblock {Setting of the magnetic structure of chiral kagome antiferromagnets by a seeded spin-orbit torque}.
\newblock {\em Sci. Adv.} \textbf{8}, eabo5930 (2022).

\bibitem{PhysRevB.99.165117}
Noky, J., Xu, Q., Felser, C. \& Sun, Y.
\newblock Large anomalous hall and nernst effects from nodal line symmetry breaking in ${\mathrm{Fe}}_{2}\mathrm{Mn}X$ ($X$ = P, As, Sb).
\newblock {\em Phys. Rev. B} \textbf{99}, 165117 (2019).

\bibitem{apl2018}
Sung, N. H., Ronning, F., Thompson, J. D. \& Bauer, E. D.
\newblock Magnetic phase dependence of the anomalous Hall effect in Mn$_3$Sn single crystals.
\newblock {\em Appl. Phys. Lett.} \textbf{112}, 132406 (2018).

\bibitem{miwa}
Miwa, S. et al.
\newblock Giant Effective Damping of Octupole Oscillation in an Antiferromagnetic Weyl Semimetal.
\newblock {\em Small Sci.} \textbf{1}, 2000062(2021).

\bibitem{pizo}
Ikhlas, M. et al.
\newblock Piezomagnetic switching of the anomalous Hall effect in an antiferromagnet at room temperature.
\newblock {\em Nat. Phys.} \textbf{18}, 1086 (2022).

\bibitem{monk}
Monkhorst, H~J. \&  Pack, J~D.
\newblock Special points for brillouin-zone integrations.
\newblock {\em Phys. Rev. B} \textbf{13}, 5188-5192 (1976).

\bibitem{VWN}
Vosko, S.H., Wilk, L., \& Nusair, M.
\newblock Accurate spin-dependent electron liquid correlation energies for local spin density calculations: A critical analysis.
\newblock {\em Can. J. Phys.} \textbf{58}, 1200-1211 (1980).

\bibitem{wan1}
Wang, X., Yates, J~R., Souza, I. \& Vanderbilt, D.
\newblock Ab initio calculation of the anomalous hall conductivity by wannier interpolation.
\newblock {\em Phys. Rev. B} \textbf{74} 195118 (2006).

\bibitem{Freimuth-2008}
Freimuth, F., Mokrousov, Y., Wortmann, D., Heinze, S. \& Blügel, S.
\newblock Maximally localized wannier functions within the flapw formalism.
\newblock {\em Phys. Rev. B} \textbf{78}, 035120 (2008).

\bibitem{wan2}
Marzari, N.
\newblock Maximally localized wannier functions: Theory and applications.
\newblock {\em Rev. Mod. Phys.} \textbf{84}, 1419-1475, (2012).

\bibitem{Yao-2004}
Yao, Y. et al.
\newblock First principles calculation of anomalous hall conductivity in ferromagnetic bcc Fe.
\newblock {\em Phys. Rev. Lett.} \textbf{92}, 037204 (2004).



\end{thebibliography}
\end{document}